\documentclass[journal]{IEEEtran}
\usepackage{cite}
\usepackage{amsmath,amssymb,amsfonts}
\usepackage{algorithmic}
\usepackage{graphicx}
\usepackage{textcomp}
\usepackage{xcolor}
\usepackage{comment}

\usepackage{url}
\usepackage{booktabs} 
\usepackage{colortbl}
\usepackage[ruled]{algorithm2e} 

\usepackage{subcaption}
\usepackage{multirow}
\usepackage{tabularx}
 \usepackage{mathrsfs}
\usepackage{graphicx}
\usepackage{rotating}
\usepackage{balance}

\usepackage{bm}
\usepackage{scalerel}

\def\BibTeX{{\rm B\kern-.05em{\sc i\kern-.025em b}\kern-.08em
    T\kern-.1667em\lower.7ex\hbox{E}\kern-.125emX}}
\makeatletter

\def\endthebibliography{%
  \def\@noitemerr{\@latex@warning{Empty `thebibliography' environment}}%
  \endlist
}
\makeatother

\begin{document}


\title{Adversarial Representation Learning for Robust Patient-Independent Epileptic Seizure Detection}

\author{Xiang Zhang,~\IEEEmembership{Member, IEEE,}
        Lina Yao,~\IEEEmembership{Member, IEEE,}
        Manqing Dong,~\IEEEmembership{Member, IEEE,}\protect\\
        Zhe Liu,~\IEEEmembership{Student Member, IEEE,}
        Yu Zhang,~\IEEEmembership{Senior Member, IEEE,}
        and~Yong Li,~\IEEEmembership{Member, IEEE}
\thanks{Xiang Zhang, Lina Yao, Manqing Dong, Zhe Liu are with the School of Computer Science and Engineering, University of New South Wales, Sydney, Australia.\protect\\
E-mail: \{xiang.zhang3, manqing.dong, zhe.liu1\}@student.unsw.edu.au, \protect\\
lina.yao@unsw.edu.au}
\thanks{Yu Zhang is with the Stanford University School of Medicine, CA, USA. E-mail: yzhangsu@stanford.edu}
\thanks{Yong Li is with the Department of Electronic Engineering, Tsinghua University, Beijing, China.
Email: liyong07@tsinghua.edu.cn}
}

\markboth{Journal of \LaTeX\ Class Files,~Vol.~14, No.~8, August~2015}%
{Shell \MakeLowercase{\textit{et al.}}: Bare Demo of IEEEtran.cls for IEEE Journals}

\maketitle

\begin{abstract}

Epilepsy is a chronic neurological disorder characterized by the occurrence of spontaneous seizures, which affects about one percent of the world’s population.
Most of the current seizure detection approaches strongly rely on patient history records and thus fail in the patient-independent situation of detecting the new patients.
To overcome such limitation, we propose a robust and explainable epileptic seizure detection model that effectively learns from seizure states while eliminates the inter-patient noises.
A complex deep neural network model is proposed to learn the pure seizure-specific representation from the raw non-invasive electroencephalography (EEG) signals through adversarial training. Furthermore, to enhance the explainability, we develop an attention mechanism to automatically learn the importance of each EEG channels in the seizure diagnosis procedure. 
The proposed approach is evaluated over the Temple University Hospital EEG (TUH EEG) database. The experimental results illustrate that our model outperforms the competitive state-of-the-art baselines with low latency. Moreover, the designed attention mechanism is demonstrated ables to provide fine-grained information for pathological analysis.
We propose an effective and efficient patient-independent diagnosis approach of epileptic seizure based on raw EEG signals without manually feature engineering, which is a step toward the development of large-scale deployment for real-life use.


\end{abstract}


\IEEEpeerreviewmaketitle

\section{Introduction} 
\label{sec:introduction} 


\IEEEPARstart{E}{pilepsy} is a chronic neurological disorder that affects about 1\% population in the world \cite{hosseini2017optimized,detti2018patient}. Such abnormal brain activity may cause seizures, unusual behavior, and sensations, and sometimes loss of awareness. 
Thus, an accurate and timely diagnosis for epileptic seizure is crucial to reduce both life and financial cost.
Non-invasive Electroencephalogram (EEG) spontaneous is the most common method for seizure diagnosis \cite{zhang2019survey} but faces several challenges.

\subsection{Challenges} 
\label{sub:challenges}

First, most of the existing epileptic seizure detection methods focus on patient-dependent scenario \cite{detti2018patient,altaf201516} but rarely consider the patient-independent situation\footnote{In some literature, the patient dependent/independent are called patient specific/non-specific.}. 
The former refers to detect the epileptic seizure of a patient by learning from his own historical records, while the latter learn from the records from other patients. 
Patient-dependent methods can achieve high accuracy because the training and testing samples are gathered from the same source and have similar distribution. Patient-dependent algorithms have been widely investigated in the last decade due to the good performance for recorded patients, however, that becomes deficient for new patients since the data distributions from different patients are much more diverse.
In contrast, patient-independent methods advance in alerting potential patients but are easily corrupted by inter-patient noises (e.g., gender, age, and epileptic type).
A majority of existing studies fail to eliminate such noises, and, to the best of our knowledge, few studies try to eliminate the inter-patient corruption by modeling the common features from the training samples. 
Thus, in this study, we propose a robust method that can learn the common seizure pattern while also mitigate the influence of inter-patient factors.

Second, the non-invasive EEG signals are with low signal-to-noise ratio and are fragile to subjective factors (e.g., emotion and fatigue) and environmental factors (e.g., noise) \cite{li2015feature,goh2018spatio}.
To discover a latent and informative representation of the signals, most traditional seizure detection methods do feature engineering on raw EEG signals, which is time-consuming and highly dependent on expertise \cite{boashash2016automatic}.
Recently, deep learning has shown great success in lots of research topics (e.g., computer vision, natural language processing, and brain-computer interface) 
because of the excellent automatic feature learning ability \cite{thodoroff2016learning}. 
Therefore, deep learning algorithms have great potential in reducing the negative effects of manual feature engineering by automatically capturing the key difference between epileptic and non-epileptic patterns.

Another issue of current epileptic seizure diagnosis approaches is about the explainability. 
Generally, we are interested in not only the seizure states of a patient but also the fine-grained observations.
For example, the epileptic seizure may occur in one specific brain region (e.g., temporal lobe) that is known as focal seizure or several brain regions (e.g., frontal, temporal, and occipital lobes) that is known as generalized epilepsy \cite{arunkumar2018entropy}. 
 The current seizure tests can provide the overall detection results while are unable to show the specific active brain regions.
To this end, for a specific patient, we attempt to design a powerful model, involving attention mechanism \cite{yuan2019fusionatt}, which can not only recognize the seizure state but also discover which EEG channels are more important to the diagnosis decision. 

\subsection{Motivation} 
\label{sub:motivation}
The key target of this work is to develop an automatic representation learning algorithm that is dependent on neurological status (e.g., seizure or normal) while is independent of person states (e.g., age and gender). The representation is learned from non-invasive EEG signals and further be used to diagnose the epileptic seizure. 
As mentioned above, a main challenge in patient-independent seizure feature learning is that the learning process can be easily affected by patient-related information (e.g., age, gender, and emotion). In order to eliminate such influences, we propose an adversarial method \cite{zhang2019adversarial} to decompose the EEG signals into a seizure-related component and a patient-related component. 

In short, the intuitive idea of this work is separating the input raw EEG signals (denoted by $\mathcal{E}$) into two parts: 
a seizure-related component, denoted by $\mathcal{S}$, that contains informative descriptions about seizure state and is insensitive to the patient identity; and a patient-related component, denoted by $\mathcal{P}$, that includes patient identity information and is insensitive to the seizure state.
The following two constraints should be satisfied during the process.
First, the sum of seizure-related component $\mathcal{S}$ and patient-related component $\mathcal{P}$ should be equal to the original EEG signal $\mathcal{E}$. 
And second, the decomposed $\mathcal{S}$ and $\mathcal{P}$ are supposed to contain pure and informative seizure and patient features, respectively. 

\subsection{Contributions} 
\label{sub:contributions}

We propose an adversarial representation learning framework to construct a robust patient-independent detection algorithm that exterminates the corruptions by inter-patient noises. The framework harnesses both deep generative model and convolutional discriminative model to capture informative seizure and patient representations.
Moreover, we improve the explainability by incorporating attention mechanism. The model can not only enhance the contribution of important EEG channels but also visualize the determine brain regions.
The main contributions of this work are highlighted as follows:

\begin{itemize} 
    \item We propose a robust framework targeting patient-independent epileptic seizure detection. The proposed approach is efficient in capturing the seizure-specific representations directly from the raw EEG signals. The resuable Python-based code can be find here\footnote{https://github.com/xiangzhang1015/adversarial\_seizure\_detection}.
    \item We propose a unified adversarial learning framework to extract the informative seizure-specific representation while eliminating the inter-patient noises. Moreover, the attention mechanism enables our approach to automatically explore the importance of each EEG channel, consequently, to bring fine-grained analysis for seizure diagnosis.
    \item We implement extensive experimental comparisons in order to evaluate the proposed approach over a benchmark seizure diagnosis dataset. The results demonstrate that our model outperforms competitive state-of-the-art methods effectively and efficiently.
\end{itemize} 



\section{Related Work} 
\label{sec:related_work}
To date, numerous analysis methods are proposed in order to automatically diagnose epileptic seizure. Non-invasive EEG signals provide important information about epileptogenic networks that must be analyzed and understood before the initiation of therapeutic procedures \cite{faust2015wavelet}. In order to discover the latent informative features, Bhattacharyya et al. \cite{bhattacharyya2017multivariate} adopted an empirical wavelet transform method to explore the multivariate signals in the time-frequency domain, in which the instantaneous amplitudes and frequencies were jointly considered. 
Moreover, Fan et al. \cite{fan2018detecting} investigated the spatial-temporal synchronization pattern in epileptic brainwaves through spectral graphical representation and developed an efficient multivariate approach for detecting seizure in real-time. Detti et al. \cite{detti2018patient} attempted to find the distinguishable non-invasive EEG patterns in order to predict the epileptic seizure onset a short-term (e.g., a few minutes) ahead. 

Apart from traditional statistic models, deep learning-based models have gained increasing attention. Schirrmeister et al. \cite{schirrmeister2017deep} adopted a deep learning-based architecture, ConvNets, to diagnose pathological patients from normal individuals based on the decoding of EEG samples. The experiments showed that the deep neural networks outperformed the conventional linear analysis classifiers. 
Lin et al. \cite{lin2016classification} proposed a deep framework for the automatic detection of epileptic EEG by combining a stacked sparse autoencoder and a softmax classifier, which firstly learned the sparse and high-level representations from the preprocessed data and then sent these representations into a softmax classifier for automatic diagnosis.

\textbf{Summary.} However, all the mentioned seizure diagnosis methods (either traditional or deep learning models) are focusing on the patient-dependent scenario which means that the training EEG samples and testing samples are from the same or same group patients. Although very limited studies mentioned patient-independent validation like \cite{orosco2016patient}, they attempt to extract the general representations through enhancing training dataset instead of designing a robust patient-independent feature learning algorithm. 
In this way, the diagnosis model can handle intra-patient factors but fail to eliminate inter-patient noise. The clinical therapy requires more complex situations (patient-independent) where the testing patient is unseen in the training stage. It is much difficult to perform accurate seizure diagnosis considering personal factors such as chronological age, gender, characteristics, and health state. In summary, proposing a robust method to deal with the patient-independent challenge is necessary.


\section{Methodology} 
\label{sec:methodology}

\begin{figure*}[!t]
\centering
  \includegraphics[width=\linewidth]{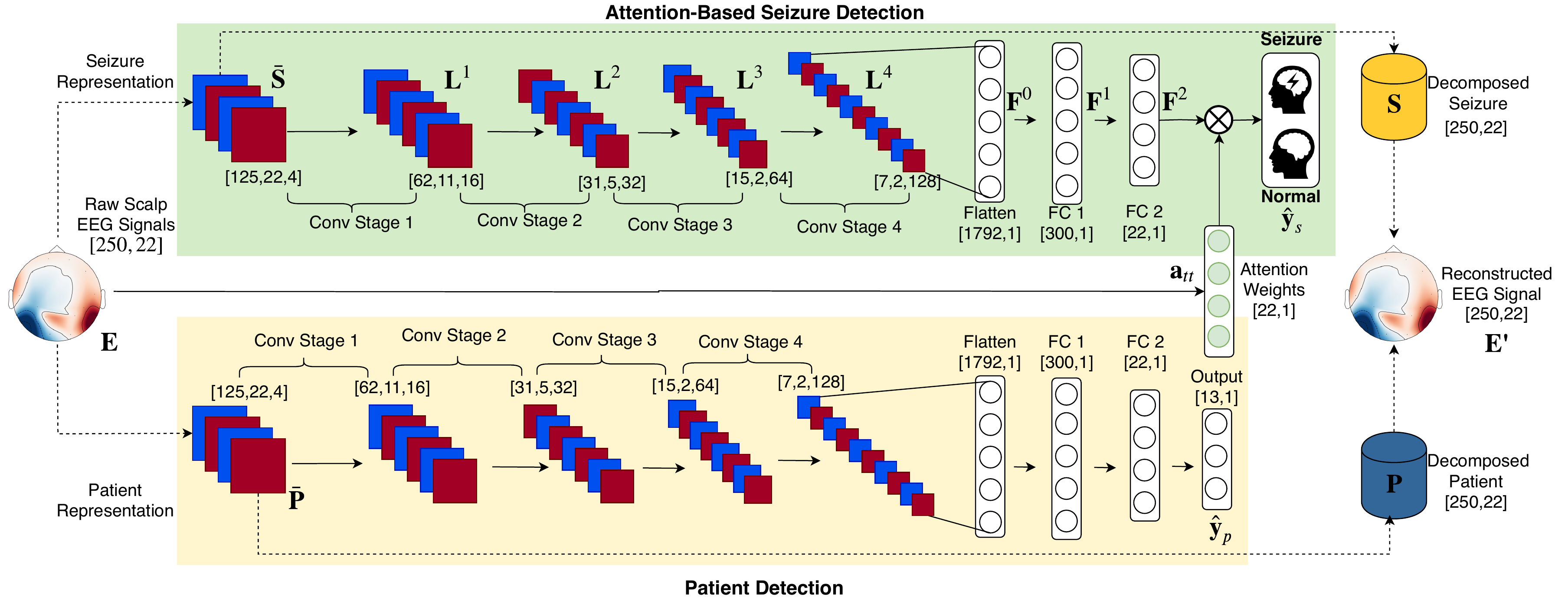}
  \caption{Workflow of the proposed patient-independent epileptic seizure diagnosis method. The collected raw EEG signals are first embedded into a latent space and be decomposed into seizure and patient representation. Afterward, the implicate representations are used to reconstruct the seizure and patient components, respectively.  At the meantime, the representations are sent to the seizure diagnosis and patient detection classifiers in order to diagnose the user's seizure stage and recognize the user's identity. The key branch is the seizure diagnosis while the patient detection branch is adopted to separate the patient-related information from the input EEG signals. The dashed lines represent the data flow of EEG decomposition.
  }
  \label{fig:method_overview}
\end{figure*}

\subsection{Overview} 
\label{sub:overview}

The proposed patient-independent seizure representation learning method contains three sections (Figure~\ref{fig:method_overview}): the EEG decomposition, the seizure diagnosis, and the patient detection. These three sections are jointly trained in an adversarial way. To satisfy the first constraint, the EEG signals are mapped into a latent space and decompose into the seizure and patient representation corresponding to the seizure state and personal information, respectively. Then, the two latent representations are used to reconstruct the original EEG signal. If the reconstructed signal equal to the input signal, we regard the learned seizure and patient representation as effective features of the decomposed EEG data. 

In addition, to satisfy the second constraint that guarantees the purification of the decomposed representations, we design a seizure-based diagnosis section and a patient classifier section following the latent representations.
The seizure diagnosis section, which is following the seizure representation, aims to recognize the epileptic seizure from a normal state. And a satisfactory seizure representation is learned until the diagnosis achieves competitive performance.
Similarly, the patient classifier that follows the patient representation targets recognizing the user's identity, and a well-performed diagnosis assures the pureness of the patient representation.

The EEG decomposition section focuses more on the reconstruction performance of the decomposed components than the pureness of the components; conversely, the seizure diagnosis section and patient classifier consider the pureness of the two components other than the effectiveness of the representations. In a whole, the two aspects are optimized toward opposite directions, which form an adversarial training situation.
It should be noted that the patient classifier is only adopted to improve the adversarial training and help to have a better signal decomposition performance. 

Furthermore, we combine an attention mechanism with our seizure diagnosis section in order to pay more attention to the important EEG electrodes. The effectiveness of attention mechanism has been demonstrated by a number of research topics like natural language process \cite{zhou2016attention}. The modified attention-based seizure diagnosis is supposed to discover which electrodes make more contributions to the
seizure. 

\subsection{EEG Decomposition} 
\label{sub:signal_decomposition}
Assume the gathered EEG signals are denoted by $\mathcal{E} = \{\bm{E_i} \in \mathbb{R}^{M \times N} \}$, in which each EEG sample $\bm{E_i}$ has $M$ time series and $N$ channels. We suppose the EEG signals can be decomposed into the seizure-related component $\mathcal{S} = \{\bm{S_i} \in \mathbb{R}^{M \times N} \}$ and the patient-related component $\mathcal{P} = \{\bm{P_i} \in \mathbb{R}^{M \times N} \}$, described as\footnote{For simplify, we ignore the sub script.}
\begin{equation}
    \bm{E} = \bm{S} + \bm{P}
\end{equation}

Nevertheless, the $\mathcal{S}$ and $\mathcal{P}$ cannot be directly calculated through traditional methods. 
To overstep this challenge, we proposed a novel decomposition model based on deep neural networks. In particular, the $\bm{E}$ is transformed into a latent space through convolutional operation followed by a max-pooling layer:
\begin{equation}
    \bm{\bar{S}} = ReLU(\bm{w_s} \circledast \bm{E} + \bm{b_s})
\end{equation}
\begin{equation}
    \bm{\bar{P}} = ReLU(\bm{w_p} \circledast \bm{E} + \bm{b_p})
\end{equation}
where $\bm{\bar{S}}, \bm{\bar{P}} \in \mathbb{R}^{J \times K \times H}$ denote the latent seizure and patient representation, in which $J$ and $K$ denote the rows and columns of the learned representation, respectively, along with $H$ denotes the number of convolutional filters. The $\circledast$ denotes convolution operation and $\bm{w_s}, \bm{b_s}, \bm{w_p}, \bm{b_p}$ represent the corresponding weights and biases. In each convolutional layer, the activation function is ReLU and the padding method is 'SAME'. 

Then, in order to ensure the representations to contain enough discriminative information, we attempt to reconstruct the EEG signals through deconvolutional operations:
\begin{equation}
    \bm{S} = ReLU(\bm{w'_s} \circledast \bm{\bar{S}} + \bm{b'_s})
\end{equation}
\begin{equation}
    \bm{P} = ReLU(\bm{w'_p} \circledast \bm{\bar{P}} + \bm{b'_p})
\end{equation}
where $\bm{S}$ and $\bm{P}$ represent the decomposed seizure- and patient-related components, which have the same shape with the input EEG signal $\bm{E}$. 
The reconstructed signal $\bm{E'}\in \mathbb{R}^{M \times N}$ can be calculated by combining the two decomposed signals 
\begin{equation}\label{eq:w1}
    \bm{E'} = w_1\bm{S} + w_2\bm{P}; w_1+w_2 =1
\end{equation}
where $w_1$ and $w_2$ are weights.

To guarantee the decomposition performance and separate the useful seizure-related information while eliminating the intra-patient corruption, we force the reconstructed signal $\bm{E'}$ to approximate the original signal $\bm{E}$ by minimizing the Mean Square Error (MSE, a.k.a. $\ell_2$) loss between them. The MSE distance has been widely applied to design the reconstruction loss function due to its simplicity and efficiency \cite{amaral2013using}. The MSE loss indeed outperforms the cross-entropy and integral probability metrics (e.g., Wasserstein loss \cite{arjovsky2017wasserstein}) in our preliminary experiments. The MSE loss function can be calculated as
\begin{equation}
    \mathcal{L}_{D} = \left \|  \bm{E} - \bm{E'}\right \|_2
\end{equation}

The learned latent seizure $\bm{\bar{S}}$ and patient representation $\bm{\bar{P}}$, compared to the decomposed components $\bm{S}$ and $\bm{P}$, have lower dimension and represent high-level features. 
Thus, the learned representations are used for the corresponding seizure diagnosis and patient detection in the next process. 

\subsection{Attention-based Seizure Diagnosis} 
\label{sub:task_specific_classifier}

The weights of different EEG electrodes are varying in seizure diagnosis \cite{temko2011instantaneous}; thus, we involve the attention mechanism to learn the channel importance and pay different attention to various signal channels. The attention mechanism, as mentioned above, allows modeling of dependencies among input sequences \cite{vaswani2017attention} and has shown success in some research topics \cite{zhou2016attention}. Meanwhile, the excellent latent feature learning ability of Convolutional Neural Networks (CNNs) has been widely used in research areas such as computer vision \cite{zhang2017learning} and natural language process \cite{zhang2016mgnc}. 
Thus, we propose an attention-based CNN algorithm to automatically extract the distinctive information from the received seizure representation $\bm{\bar{S}}$ which has shape $[J, K, H]$. 

As shown in Figure~\ref{fig:method_overview}, the attention-based seizure diagnosis contains four convolutional stages: the flatten layer, two fully-connected layers (FC) and an output layer. Each convolutional stage contains a convolutional layer and a max-pooling layer. We assume the input layer receives the learned seizure feature $\bm{\bar{S}}$ and sent to the convolutional layers:
\begin{equation}
    \bm{C^i} = ReLU(\bm{w^i_s} \circledast \bm{L^{i-1}} + \bm{b^i_s}), i \in \{1, 2, 3\}
\end{equation}
where $\bm{C^i}$ and $\bm{L^{i}}$ denote the $i$-th convolutional and pooling layer, respectively.
If $i=1$, $\bm{L^0}$ equals to the input $\bm{\bar{S}}$; otherwise, the pooling layer can be measured by:
\begin{equation}
    \bm{L^i_u} = \underset{u \in \mathcal{U}}{max}\{\bm{C^i_u}\}    , i \in \{1, 2, 3\}
\end{equation}
where $\mathcal{U}$ represents the max-pooling perception field and $u$ denotes the $u$-th element. 

The forth pooling layer $\bm{L^4}$ is flattened to 1-D vector $\bm{F^0}$ and then fed into the FC layers:
\begin{equation}
    \bm{F^i} = sigmoid(\bm{w^i_f} \bm{F^{i-1}} + \bm{b^i_f}), f \in \{ 1, 2, 3\}
\end{equation}
where $sigmoid$ denotes the activation function and $\bm{F^i}$ denotes the $i$-th FC layer. 

The predicted seizure state $\bm{\hat{y_s}}$ is related to the last FC layer $\bm{F^2}$ and the learned attention weights $\bm{a_{tt}}$:
\begin{equation}
    \bm{\hat{y_s}} = \bm{F^2} \cdot \bm{a_{tt}}
\end{equation}
where $\cdot$ denotes dot production and $\bm{a_{tt}}$ is directly learn from the EEG sample:
\begin{equation}
\label{eq:att}
    \bm{a_{tt}} = sigmoid(\bm{w_a} \bm{E} + \bm{b_a})
\end{equation}
where $\bm{w_a}$ and $\bm{b_a}$ denote the corresponding parameters.

The cross-entropy loss function of seizure diagnosis $\mathcal{L}_s$ is defined as:
\begin{equation}
    \mathcal{L}_s = -(\bm{y_s}log(\bm{p(\bm{\hat{y_s}})})+ (1-\bm{y_s})log(1-\bm{p(\bm{\hat{y_s}})}))
\end{equation}
where $\bm{y_s}$ and $\bm{p(\bm{\hat{y_s}})}$ denote the ground truth of seizure state and the predicted probability of the patient in seizure state, respectively. 

\subsection{Patient detection} 
\label{sub:subject_detection}
The architecture of the patient detection is almost identical to the attention-based seizure diagnosis except two main differences: 1) the patient detection component does not need the attention module; 2) the patient detection performs multi-class instead of binary classification.
Similar to $\mathcal{L}_s$, we can measure the multi-class cross-entropy loss function $\mathcal{L}_p$ of the patient identity detection classifier as:
\begin{equation}
    \mathcal{L}_p = - \sum_{c=1}^{C}  (\bm{y_{p}}log(\bm{\hat{y_{p}}}))
\end{equation}
where $\bm{y_p}$, $\bm{\hat{y_{p}}}, $and $C$ denote the ground truth of patient identity, the predicted identity, and the overall number of training patients, respectively. 

\subsection{Training Details} 
\label{sub:loss_function}
In the proposed approach, we have three loss functions which are the reconstruction loss $\mathcal{L}_{D}$ in signal decomposition and the classification loss $\mathcal{L}_s$ and $\mathcal{L}_p$ in the attention-based seizure diagnosis and patient detection, respectively. 

We propose an adversarial training strategy to jointly train all the loss functions:
\begin{equation}
    \mathcal{L} = \mathcal{L}_{D} + \mathcal{L}_s + \mathcal{L}_p +\ell_2 
\end{equation}
which 
assures all losses work on the gradient at the same time and converge to a trade-off position of that balances the reconstruction performance and the purification of the decomposed signals. The $\ell_2$ represents the $\ell_2$ norm with coefficient of $0.0001$ to prevent overfitting.
Then, since the seizure diagnosis is the most crucial component in this approach, we train the $\mathcal{L}_s$ one more time in order to raise its priority. To sum up, in each training epoch, the $\mathcal{L}$ and $\mathcal{L}_s$ are optimized once in turn. We adopt the Adam optimizer with learning rate $e^{-4}$ for both loss functions. The algorithm is trained for $250$ epochs. A dropout layer with $0.8$ keep rate is added to the flatten layer in both classifiers in order to prevent overfitting. The overall flowchart is presented in algorithm~\ref{alg:method}.

\begin{algorithm}[!t]
\begin{small}
\caption{The Proposed Approach}
\label{alg:method}
\begin{algorithmic}[1]
\renewcommand{\algorithmicrequire}{\textbf{Input:}}
\renewcommand{\algorithmicensure}{\textbf{Output:}}
 \REQUIRE Raw EEG data $\bm{E}$
 \ENSURE  Seizure State: $\bm{\hat{y_s}}$
 \FOR{$\bm{E} \in \mathcal{E}$}
    \STATE \#EEG Decomposition:
    \STATE $\bm{\bar{S}} = ReLU(\bm{w_s} \circledast \bm{E} + \bm{b_s})$
    \STATE $\bm{\bar{P}} = ReLU(\bm{w_p} \circledast \bm{E} + \bm{b_p})$
    \STATE $\bm{S} = ReLU(\bm{w'_s} \circledast \bm{\bar{S}} + \bm{b'_s})$
    \STATE $\bm{P} = ReLU(\bm{w'_p} \circledast \bm{\bar{P}} + \bm{b'_p})$
    \STATE $\bm{E'} = (\bm{S} + \bm{P})/2$
    \STATE $\mathcal{L}_{D} = \left \|  \bm{E} - \bm{E'}\right \|_2$
    \STATE \#Attention-based Seizure Diagnosis:
        \STATE $\bm{F^2} \leftarrow \bm{\bar{S}}$
        \STATE $\bm{a_{tt}} = sigmoid(\bm{w_a} \bm{E} + \bm{b_a})$
        \STATE $\bm{\hat{y_s}} = \bm{F^2} \cdot \bm{a_{tt}}$
        \STATE $\mathcal{L}_s \leftarrow \bm{y_s}, \bm{\hat{y_s}}$
    \STATE \#Patient Detection:
        \STATE $\bm{\hat{y_p}} \leftarrow \bm{\bar{P}}$
        \STATE $\mathcal{L}_p \leftarrow \bm{y_p}, \bm{\hat{y_p}}$
\ENDFOR
\STATE Minimize $\mathcal{L} = \mathcal{L}_{D} + \mathcal{L}_s + \mathcal{L}_p +\ell_2$
\RETURN $\bm{\hat{y_s}}$
\end{algorithmic}
\end{small}
\end{algorithm}

\section{Experiments} 
\label{sec:experiments}
In this section, we evaluate the proposed approach over a benchmark dataset of an epileptic seizure. First of all, we present the experimental setting and hyper-parameters in the training model. Then we report the performance comparison results among our method with several competitive state-of-the-art baselines. The comparison is assessed by metrics (such as accuracy, sensitivity, specificity, ROC curves, and AUC scores). Moreover, we provide the latency analysis to demonstrate the efficiency of our method. At last, we attempt to explore the importance of each brain region in seizure diagnosis and present our empirical hypothesis.

\subsection{Dataset} 
\label{sub:dataset}
The proposed approach is evaluated over a benchmark dataset TUH corpus \cite{obeid2016temple} for patient-independent epileptic seizure diagnosis. The TUH is a neurological seizure dataset of clinical EEG recordings associated with $22$ channels from a $10/20$ configuration. Since the epileptic seizure may only occur on several channels, we define the seizure onset when more than $12$ channels having a seizure. In order to have enough seizure samples, we only select the subject who contains more than $250$ seconds of seizure state. There are 14 subjects meet the criteria.
For each subject, we use $500$ seconds (half normal and half seizure) of EEG signals with the sampling rate of $250$ Hz. Each EEG sample has a window size of $250$ (i.e., lasting for 1 second) with $50\%$ overlapping. There are 13,986 samples in total. The epileptic seizure state is labelled as 1 while the normal state is labelled as 0.
In the evaluation, we adopt Leave-One-Out strategy, i.e., 14-fold cross-validation, which is one subject work as the testing set while all the other subjects are used for training. For example, the accuracy of subject 0 is calculated when subjects 1 $\sim$ 13 are used for training while subject 0 is used for testing. 

\begin{table*}[t]
\centering
\caption{The accuracy comparison among our methods with the baselines and state-of-the-art studies. We adopt Leave-One-Out strategy for patient-independent testing. }
\label{tab:comparison}
\resizebox{\textwidth}{!}{
\begin{tabular}{llllllllllllllll} 
\hline
\multirow{2}{*}{\textbf{Methods}} & \multicolumn{15}{c}{\textbf{Subject ID}} \\ \cline{2-16}
 & \textbf{0} & \textbf{1} & \textbf{2} & \textbf{3} & \textbf{4} & \textbf{5} & \textbf{6} & \textbf{7} & \textbf{8} & \textbf{9} & \textbf{10} & \textbf{11} & \textbf{12} & \textbf{13} & \textbf{Average} \\ \hline
\textbf{SVM} & 0.786 & 0.358 & 0.842 & 0.684 & 0.67 & 0.61 & 0.824 & 0.868 & 0.252 & 0.558 & 0.898 & 0.4 & 0.614 & 0.528 & 0.643 \\
\textbf{RF} & 0.398 & 0.681 & 0.92 & 0.754 & 0.314 & 0.634 & 0.696 & 0.568 & 0.298 & 0.602 & 0.906 & 0.612 & 0.658 & 0.603 & 0.619 \\
\textbf{KNN} & 0.786 & 0.652 & 0.878 & 0.674 & 0.682 & 0.608 & 0.808 & 0.618 & 0.562 & 0.612 & 0.882 & \textbf{0.652} & 0.672 & 0.647 & 0.699 \\
\textbf{Schirrmeister \cite{schirrmeister2017deep}} & 0.793 & 0.743 & 0.965 & 0.758 & 0.789 & 0.665 & 0.813 & 0.871 & 0.619 & 0.634 & 0.919 & 0.571 & 0.744 & 0.711 & 0.760 \\
\textbf{Ansari \cite{ansari2019neonatal}} & 0.739 & 0.757 & 0.953 & 0.749 & 0.746 & 0.677 & 0.807 & 0.729 & 0.579 & 0.611 & 0.902 & 0.559 & 0.75 & 0.729 & 0.735 \\
\textbf{Lin \cite{lin2016classification}} & 0.848 & 0.601 & 0.832 & 0.654 & 0.832 & 0.603 & 0.802 & 0.868 & 0.668 & 0.611 & 0.876 & 0.603 & \textbf{0.784} & 0.714 & 0.736 \\
\textbf{Kiral \cite{kiral2018epileptic}} & 0.805 & 0.669 & 0.855 & 0.709 & 0.772 & 0.619 & 0.823 & 0.836 & \textbf{0.746} & 0.598 & 0.835 & 0.556 & 0.745 & 0.726 & 0.736 \\
\textbf{Ours} & \textbf{0.841} & \textbf{0.826} & \textbf{0.978} & \textbf{0.774} & \textbf{0.842} & \textbf{0.733} & \textbf{0.911} & \textbf{0.914} & 0.697 & \textbf{0.652} & \textbf{0.923} & 0.604 & 0.772 & \textbf{0.787} & \textbf{0.805} \\ \hline
\end{tabular}
}
\end{table*}

\subsection{Hyper-parameter Setting} 
\label{sub:hyper_parameter_setting}
Next, we report the hyper-parameter settings in detail. In the signal decomposition, the input EEG sample has shape $[M=250, N=22]$, the convolutional layer has 4 filters with size $[3, 3]$, and $[2, 2]$ strides. The followed max-pooling has $[2,1]$ window with $[2,1]$ strides. All the padding methods in this work are 'SAME'. The deconvolutional layers have 1 filter and the other setting are the same. 
The seizure- and patient- branches have identical hyper-parameters. 

Based on our empirical hyper-parameter tuning, the attention-based seizure diagnosis and patient detection have same hyper-parameter settings: the four convolutional layers have 16, 32, 64, 128 filters with $[3,3], [3,3], [2,2], [2,2]$ sizes (all have $[1,1]$ strides), respectively. All the kernel sizes and strides of the first three max-pooling layers are $[2,2]$ and $[2,1]$ for the last max-pooling layer. The two FC layers have $300$ and $22$ hidden neurons, respectively. The attention layer has $22$ hidden neurons corresponding to $22$ input channels. The detailed information can be found in Figure~\ref{fig:method_overview}. We also investigated the weights in Equation~\ref{eq:w1}, which show that the performance is not sensitive to the averaging weights. The experimental performance ranges in $[0.784, 0.806]$ while $w_1$ changes from 0.1 to 0.9 (while $w_2$ varies from 0.9 to 0.1).  One possible reason is that other parameters within the de/convolutional operation will automatically adjust based on the change of $w_1$, which alleviated the influence brought by $w_1$. For simplicity, we set $w_1=w_2=0.5$ in this work.

\begin{figure}[t]
    \centering
    \begin{subfigure}[t]{0.23\textwidth}
        \centering
        \includegraphics[width=\textwidth]{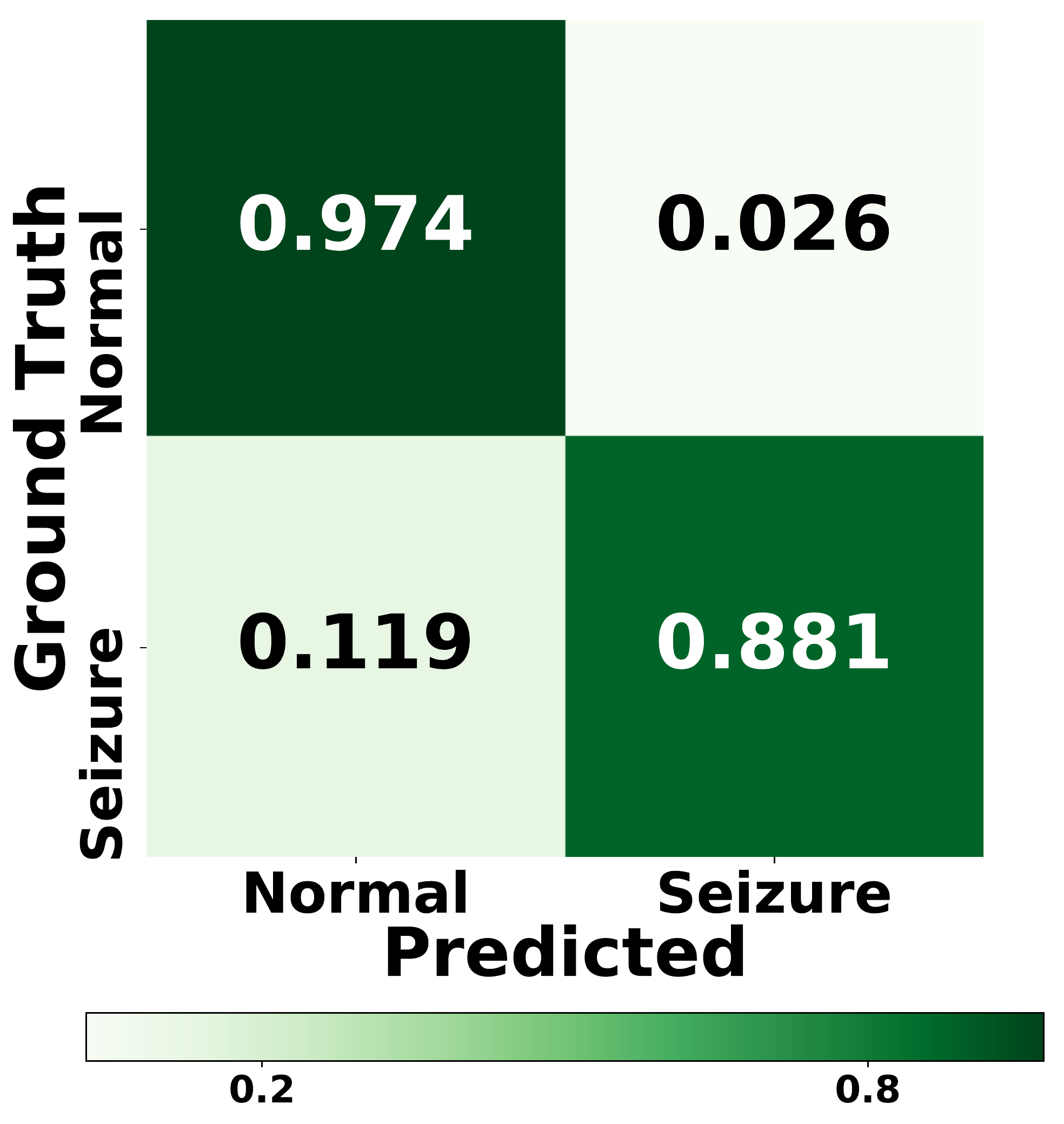}
        \caption{Confusion matrix}
        \label{fig:cm}
    \end{subfigure}%
    \hfill
    \begin{subfigure}[t]{0.25\textwidth}
        \centering
        \includegraphics[width=\textwidth]{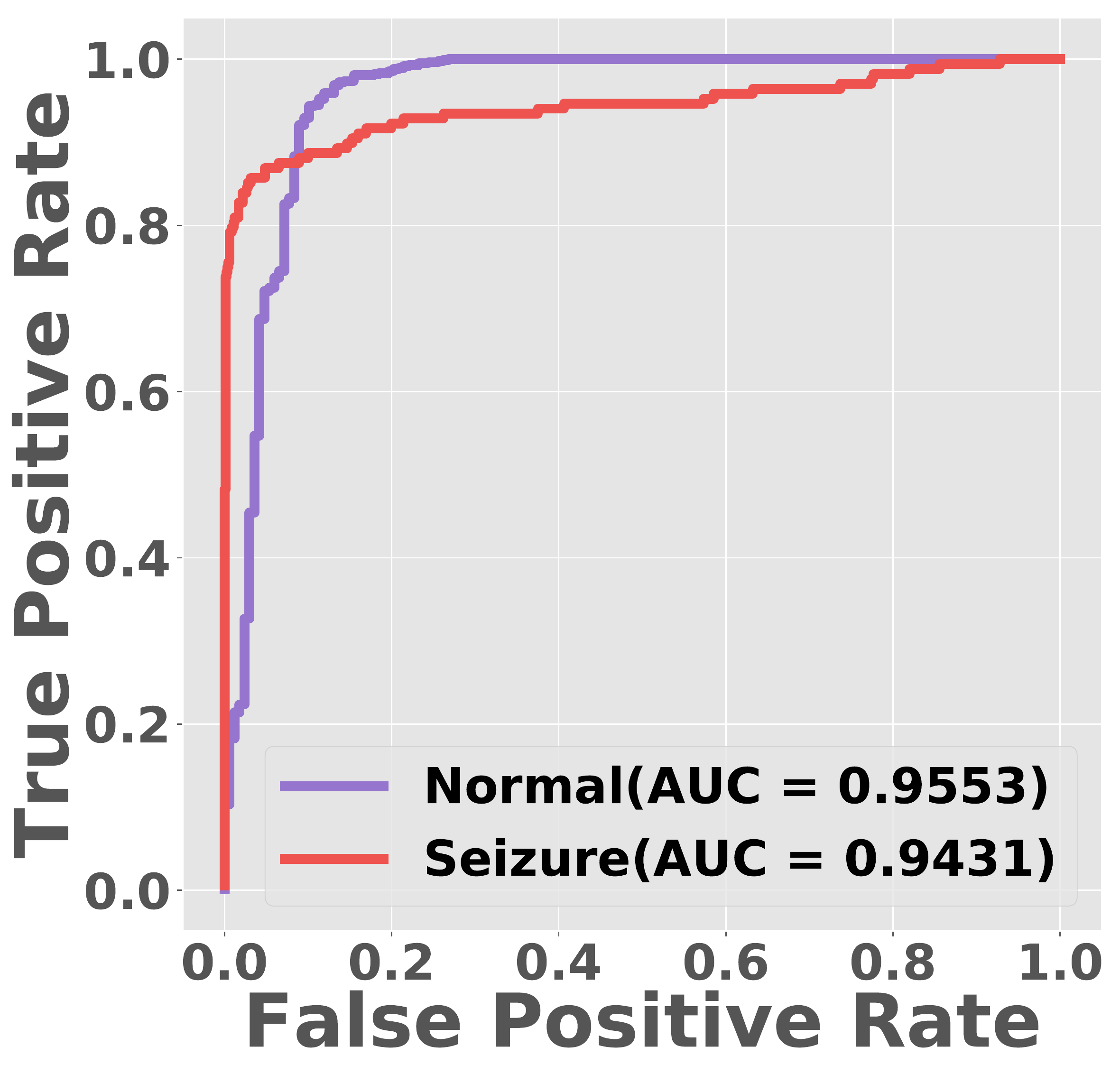}
        \caption{ROC curves with AUC}
        \label{fig:roc}
    \end{subfigure}%
    \caption{Confusion matrix and ROC curves}
    \label{fig:cm_roc}
\end{figure}

\subsection{Overall Comparison} 
\label{sub:overall_comparison}
We compare our approach with three widely-used baselines and a set of competitive state-of-the-art methods. The key parameters of the baselines are listed here: Linear SVM ($C = 1$), Random Forest (RF, $n = 50$), and K-nearest Neighours ($k = 3$)\footnote{We directly present the optimal hyper-parameter setting and omitted the tuning procedure to save space.}. 
Moreover, the comparable state-of-the-art studies for patient-independent seizure diagnosis are listed as follows. In this work, we mainly compare with the deep learning-based advance methods since it has been demonstrated that deep neural networks generally perform better than traditional models (Section~\ref{sec:related_work}).
\begin{itemize}
    \item Schirrmeister et al. \cite{schirrmeister2017deep} apply convolutional neural networks to distinguish seizure segments by decoding task-related information from EEG signals.
    \item Ansari et al. \cite{ansari2019neonatal} adopt CNN to extract the latent features which are fed into a Random Forest (RF) classifier for the final epileptic seizure detection in neonatal babies.
    \item Lin et al. \cite{lin2016classification} propose a sparse deep autoencoder with three hidden layers to extract the representative features from epileptic EEG signals.
    \item Kiral et al. \cite{kiral2018epileptic} design a deep neural network for seizure diagnosis and further develop a prediction system on a wearable device.
\end{itemize}
Since the state-of-the-art algorithms are designed and evaluated in the context of patient-dependent situations, to make them comparable, we reproduce all the state-of-the-art models in python code in our work to perform a patient-independent task. Take \cite{schirrmeister2017deep} as an example, we built a CNN model with 7 convolutional layers (the architecture and hyper-parameters are according to Table 2 in \cite{schirrmeister2017deep}) to perform patient-independent epileptic seizure detection. 
All the baselines and our method are evaluated over the same dataset with the Leave-One-Out strategy. The overall performance of all the compared methods is reported in Table~\ref{tab:comparison}. From this table, we can clearly observe that our approach outperforms all the baselines and state-of-the-art models by achieving the average accuracy of 80.5\%, illustrating the effectiveness of the proposed method in patient-independent seizure representation learning. The advantage is demonstrated under all of the 14 subjects. 

Moreover, to have a closer observation of the diagnosis results, we report the confusion matrix and receiver operating characteristic (ROC) curves with area under the curve (AUC) scores. The results of one of the best-performed subjects are shown in Figure~\ref{fig:cm_roc}. From the confusion matrix, we can observe that our approach achieves a high sensitivity of 97.4\% and a slightly low but still competitive specificity of 88.1\%. 
Moreover, the ROC curves and AUC score are presented in Figure~\ref{fig:roc}.

\begin{figure}[t]
    \centering
    \begin{subfigure}[t]{0.24\textwidth}
        \centering
        \includegraphics[width=\textwidth]{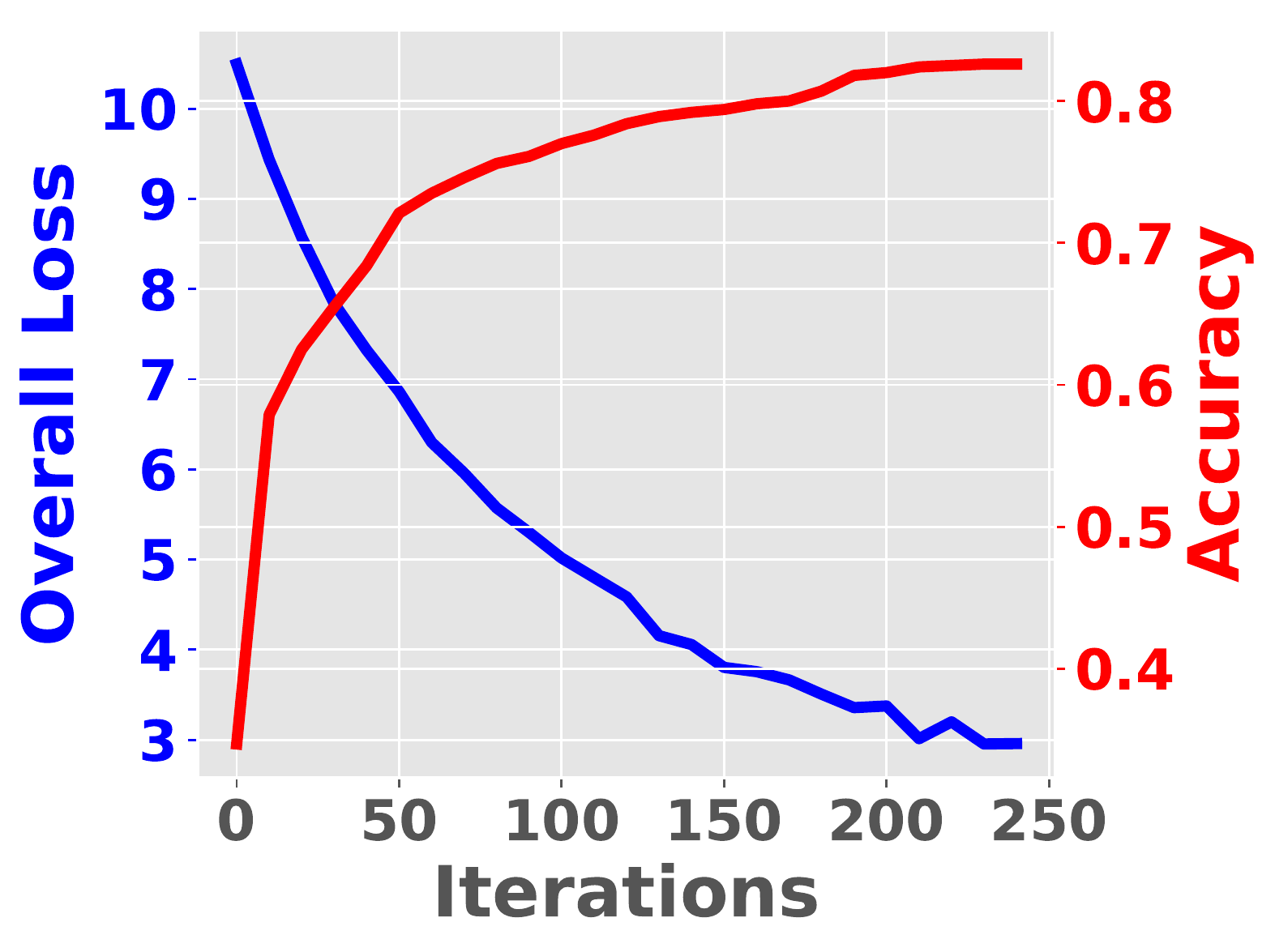}
        \caption{Overall Loss and Accuracy}
        \label{fig:loss_acc}
    \end{subfigure}%
        \hfill
    \begin{subfigure}[t]{0.24\textwidth}
        \centering
        \includegraphics[width=\textwidth]{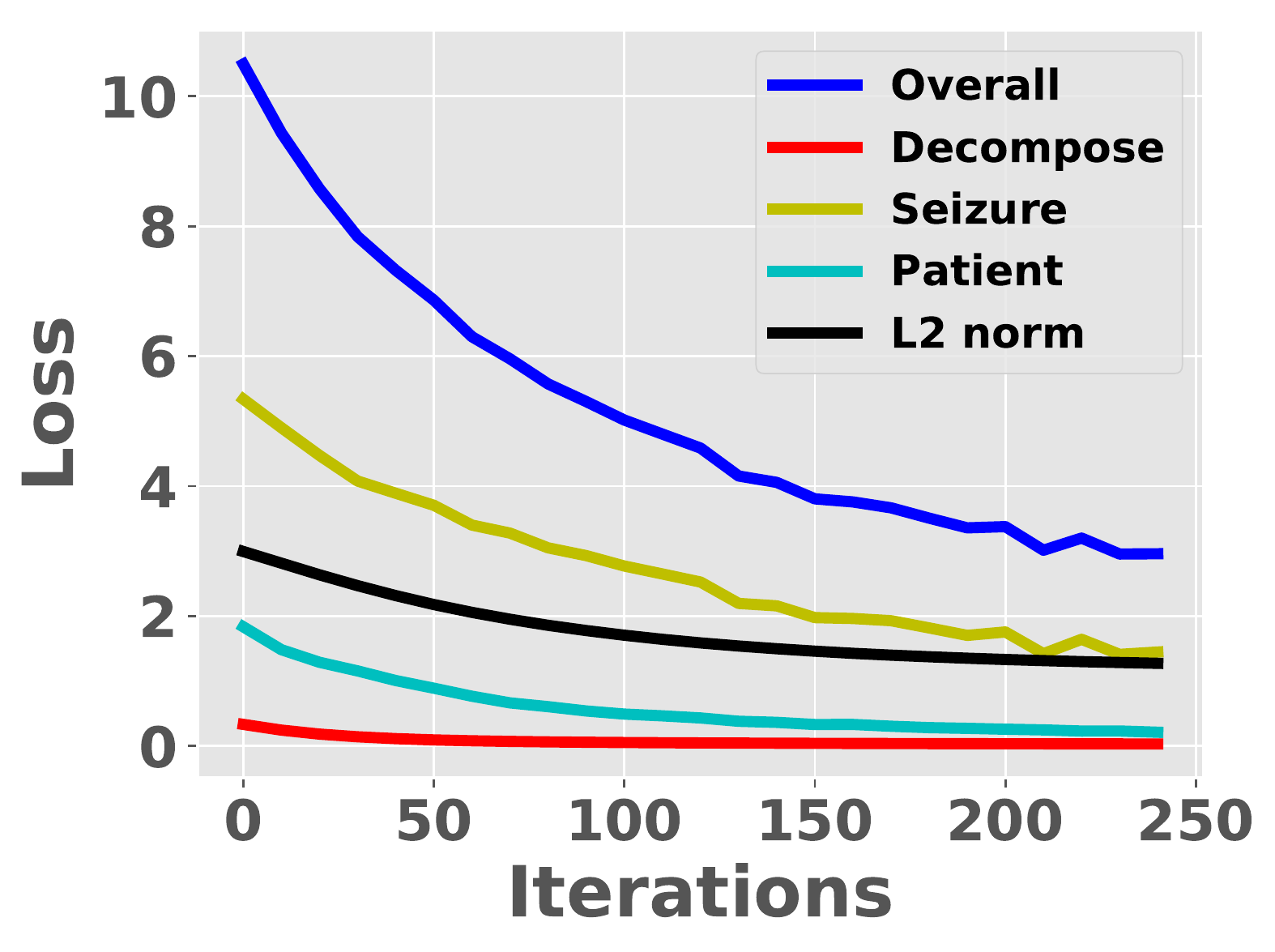}
        \caption{Loss components}
        \label{fig:loss}
    \end{subfigure}    
    \caption{Convergence trend with the increase in the number of training iterations}
    \label{fig:}
\end{figure}

Furthermore, we present the converging trend with the increase of training iterations. Figure~\ref{fig:loss_acc} tracks the evolution of the overall loss function and the testing accuracy, which shows that our model can converge smoothly. In Figure~\ref{fig:loss}, we dig deeper to investigate the trend of each loss function component. It can be observed that the seizure detection loss $\mathcal{L}_s$ is the most important loss which takes about half proportion. This also demonstrates that our strategy of training $\mathcal{L}_s$ one more time is appropriate.

\subsection{Latency Analysis} 
\label{sub:latency_analysis}
In addition to the diagnosis performance, latency is another important metric for an EEG-based epileptic seizure diagnosis system. 
In this section, we present the latency comparison among the proposed method and the state-of-the-art baselines. All the experiments are run in an NVIDIA TITAN X GPU platform with 10 Gbps memory speed and 3584 cores. 

We evaluate the training time and testing latency of our method. The results show that we only requires $0.06$s ($0.01$s for EEG decomposition while $0.05s$ for attention-based seizure diagnosis) for testing although $1417.4$s for training.
In summary, the diagnosis latency of our method is acceptable in a potential online epileptic seizure diagnosis system.

\begin{figure*}[t]
    \centering
    \begin{subfigure}[t]{0.24\textwidth}
        \centering
        \includegraphics[width=\textwidth]{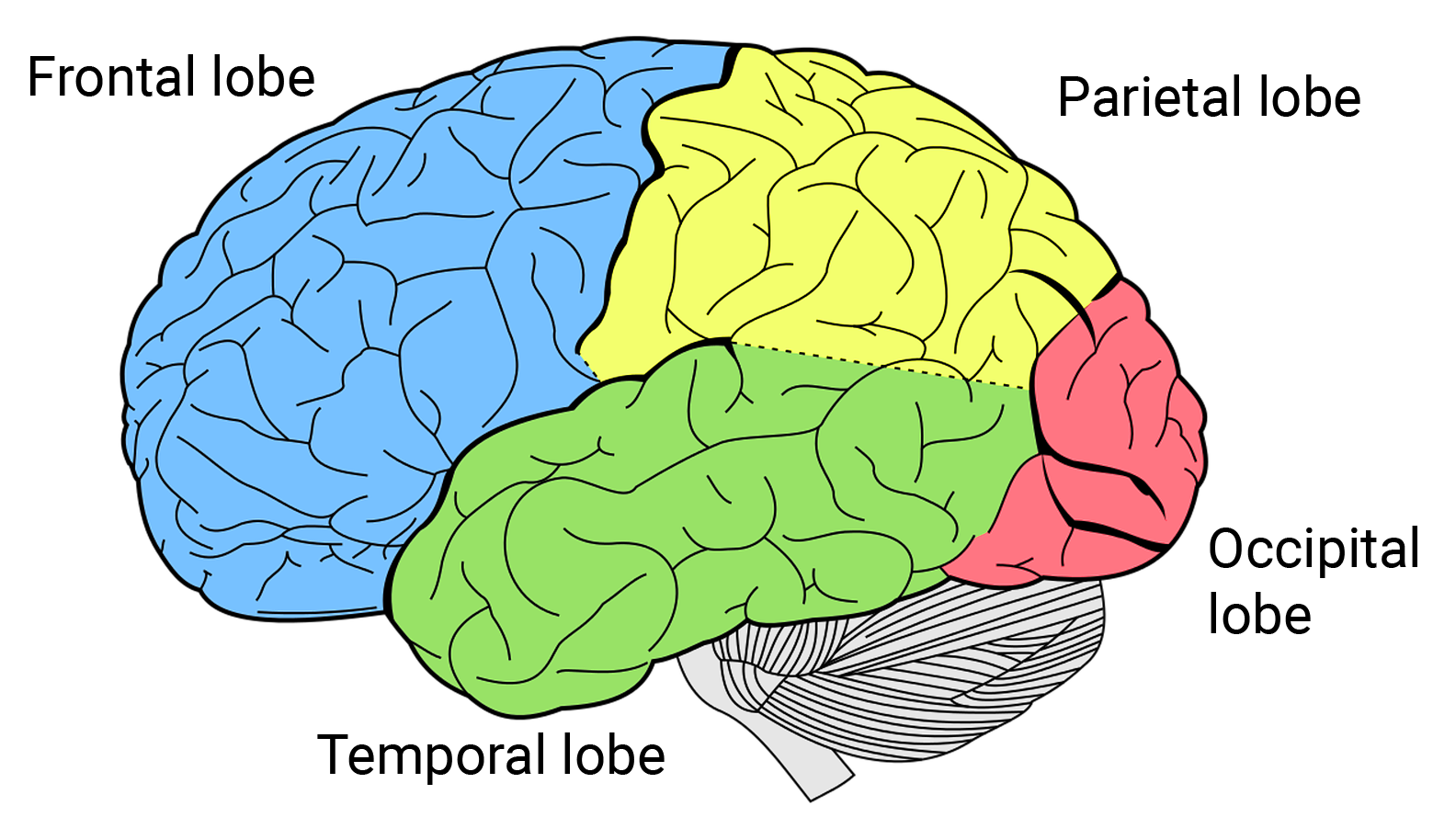}
        \caption{Brain lobes distribution}
        \label{fig:lobes}
    \end{subfigure}%
    \begin{subfigure}[t]{0.24\textwidth}
        \centering
        \includegraphics[width=0.8\textwidth]{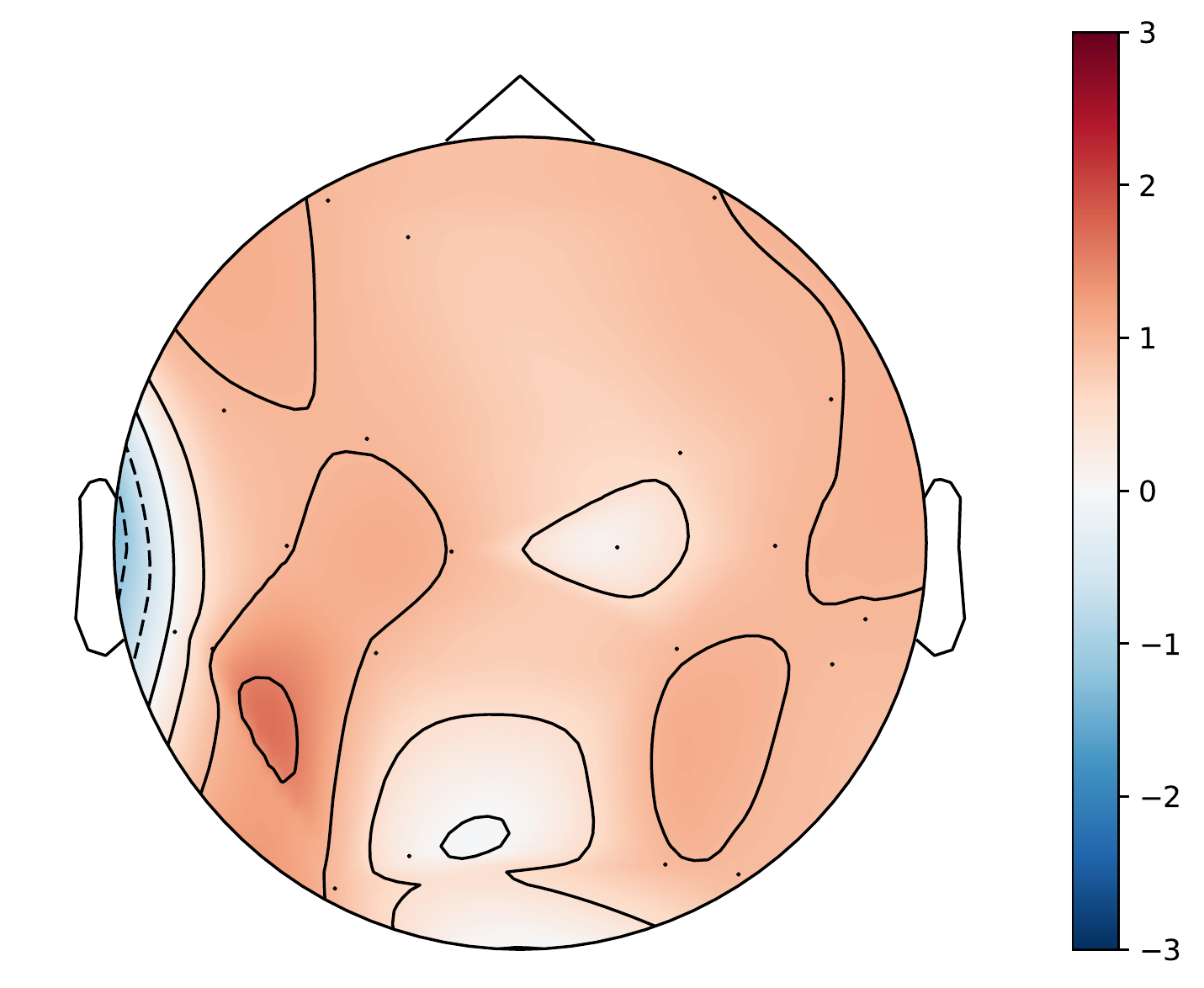}
        \caption{Attention topography (S1)}
        \label{fig:attention_1}
    \end{subfigure}    
    \begin{subfigure}[t]{0.24\textwidth}
    \centering
    \includegraphics[width=0.8\textwidth]{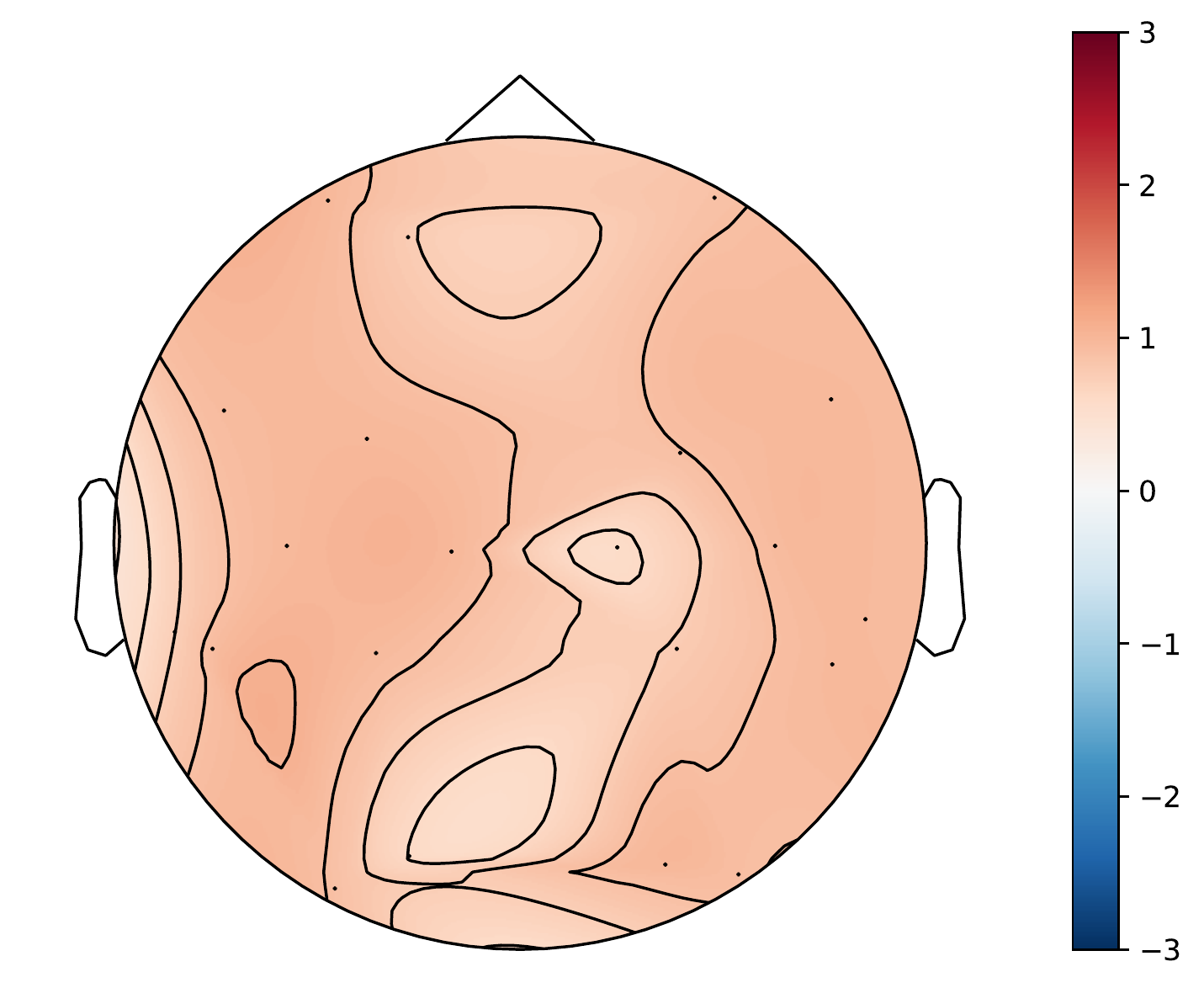}
    \caption{Attention topography (S8)}
    \label{fig:attention_2}
    \end{subfigure}   
    \begin{subfigure}[t]{0.24\textwidth}
    \centering
    \includegraphics[width=0.8\textwidth]{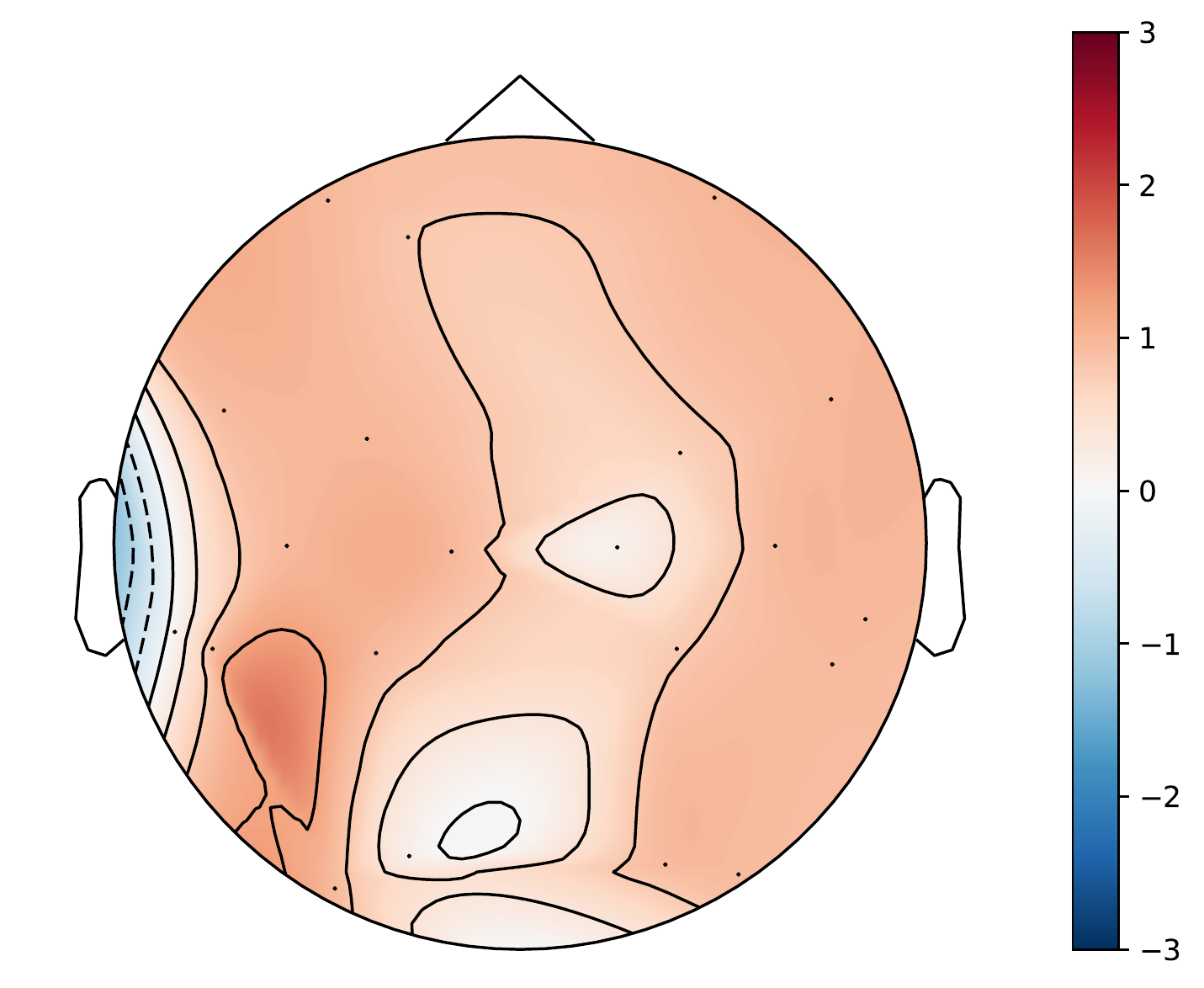}
    \caption{Attention topography (S12)}
    \label{fig:attention_3}
    \end{subfigure}   
    \caption{Demonstration of the learned channel attention topography of three randomly selected subjects (S1, S8, S12)
    . It can be inferred from the topography that the epileptic seizure occurs from the temporal lobe around the T5 channel and then spread to all the brain surfaces except the earlobes.
    }
    \label{fig:topography}
\end{figure*}

\subsection{Explainability Analysis} 
\label{sub:attention}
Explainability is important to a deep learning framework in order to have a better understanding of how the system works. Our method is enabled to capture the importance of each brain region in the diagnosis of epileptic seizure. The proposed method is designed to automatically pay different attention to various EEG channels. For a well-trained and effective model, the learned attention weights can be regarded as the importance of the channels. We conduct leave-one-out cross-validation experiments to investigate the leaned attention topography of each subject (we only present the topographies of three randomly selected subjects due to page limitation). In this section we report which region in the brain hemisphere contributes more to the epileptic seizure.

Figure~\ref{fig:lobes} shows the diagrammatic representation of brain lobes\footnote{https://socratic.org/questions/what-lobe-of-the-brain-is-responsible-for-vision} including frontal, temporal, parietal, and occipital lobes while Figure~\ref{fig:attention_1} to Figure~\ref{fig:attention_3} illustrate the topographies of the learned channel attention from three different subjects based on Equation~\ref{eq:att}. 
The red color represents higher weight while blue color indicates lower weight. 
From the attention topographies, we can observe that the left temporal lobe (around $T5$ channel) always have the highest weights across all the subjects, which indicates the temporal lobe has a close relationship with epileptic seizure. 
 This conclusion is consistent with some biological inference that a region of particular importance in adults with epilepsy is mesial of the temporal lobe \cite{lerche2013ion}. This region is comprised of such structures as the hippocampus and amygdala, which control emotion, and the uncus, which is responsible for processing smells. This is why seizures generally bring strong emotion effects (e.g., fear) and sensation of an acrid odor (e.g., burning rubber)\footnote{https://www.massgeneral.org/childhood-epilepsy/overview/brain.aspx}.
Second, the brain regions near the left earlobe (around $A1$ channel) and the left part of the occipital lobe (near $O1$) seem have a rather lower weights indicating the less significance in seizure diagnosis.
The patterns keeps consistent across different patients, which illustrates that the proposed attention model is reasonable and effective.

The above-mentioned hypotheses are empirical inference based on the designed explainable model and we hope these results could bring some inspiration to researchers in biological, brain science, and medical areas.

\section{Discussion and Future Work} 
\label{sec:discussion_and_future_work}

In this work, we propose a novel deep learning framework aiming at patient-independent epileptic seizure diagnosis, which is demonstrated has competitive performance and high interpretability.

Our method can protect the patients' privacy. By extracting the seizure representation in the EEG decomposition part, the learned $\bar{\bm{S}}$ contains rare person-related information. Thus, it's difficult to infer the patient's identity from the $\bar{\bm{S}}$. In the real-world deployment, the patients' privacy is protected.

Next, we discuss the opening challenges of this work and the potential future work.
First, our model achieved the average patient-independent diagnosis accuracy of around 80\%, which outperforms the state-of-the-art baselines but still not enough in clinical deployment. 
One feasible solution is adopt ensembling strategies such as voting, bootstrap, aggregating. Take voting strategy as an example, we select multiple EEG segments and predict the seizure state independently. 
Then, the final decision is the votes from all the prediction results. Such strategy may increase the latency but improves the prediction accuracy. In the perspective of time efficiency, if we have five EEG segments where each segment costs one second, the data collection latency will be 5 seconds, which is acceptable in a real-world scenario. In the perspective of prediction performance, suppose the original probability of correct classification $p$ is $0.8$; then, after voting from all of the EEG segments, e.g. 5 segments, the probability becomes  $1-[C^4_5p(1-p)^4-C^5_5(1-p)^5] = 0.9933$, which is satisfactory in most of the application scenarios. Thus, we can choose an appropriate number of segments to assure the model achieve better detection accuracy while is time-efficient in real-world cases. 

Moreover, the proposed approach has a relatively high complexity containing three computational components. As a result, the training time of our model is much longer than the state-of-the-art models. One of our future scopes is to develop a lightweight framework to save computational resources. 

By the way, the proposed method contains a large set of hyper-parameters (such as the convolutional filter size) which may lead to heavy tuning workload. One possible method is to adopt some hyper-parameter optimization algorithm (e.g., orthogonal array tuning method \cite{zhang2019deep}) to efficiently learn the optimal hyper-parameter settings.

In addition, the seizure samples in the training stage are noisy because they may have different seizure EEG channels. For instance, some samples may have 12 seizure channels while some have 21 seizure samples, however, all of them are labelled as 'seizure'. This phenomenon corrupts the training data quality, disturbs the data distribution, and has a negative impact on the interpretation. Nevertheless, the dataset does not contain enough samples in which all the channels are seizure. Thus, to keep the balance between the number of seizure and normal samples, we have to set a lower threshold as `12 seizure channels' to form more seizure samples (Section~\ref{sub:dataset}).

\section{Conclusion} 
\label{sec:conclusion}
In this paper, we propose a novel and generic deep learning framework aiming at patient-independent epileptic seizure diagnosis. The proposed approach refines the seizure-specific representation by eliminating the inter-subject noise through adversarial training. Moreover, we involve the attention mechanism to learn the contribution of each EEG channel in the epileptic seizure detection, which empowers our method with great explainability. We conduct extensive experiments over a benchmark dataset. The experimental results show that the proposed approach not only outperforms a batch of state-of-the-art methods by a large margin but also indicates low testing latency and highly explainability. 

\balance
\bibliographystyle{IEEEtran}
\bibliography{JBHI_seizure.bib} 

\begin{thebibliography}{10}
\providecommand{\url}[1]{#1}
\csname url@samestyle\endcsname
\providecommand{\newblock}{\relax}
\providecommand{\bibinfo}[2]{#2}
\providecommand{\BIBentrySTDinterwordspacing}{\spaceskip=0pt\relax}
\providecommand{\BIBentryALTinterwordstretchfactor}{4}
\providecommand{\BIBentryALTinterwordspacing}{\spaceskip=\fontdimen2\font plus
\BIBentryALTinterwordstretchfactor\fontdimen3\font minus
  \fontdimen4\font\relax}
\providecommand{\BIBforeignlanguage}[2]{{%
\expandafter\ifx\csname l@#1\endcsname\relax
\typeout{** WARNING: IEEEtran.bst: No hyphenation pattern has been}%
\typeout{** loaded for the language `#1'. Using the pattern for}%
\typeout{** the default language instead.}%
\else
\language=\csname l@#1\endcsname
\fi
#2}}
\providecommand{\BIBdecl}{\relax}
\BIBdecl

\bibitem{hosseini2017optimized}
M.-P. Hosseini, D.~Pompili, K.~Elisevich, and H.~Soltanian-Zadeh, ``Optimized
  deep learning for eeg big data and seizure prediction bci via internet of
  things,'' \emph{IEEE Transactions on Big Data}, vol.~3, no.~4, pp. 392--404,
  2017.

\bibitem{detti2018patient}
P.~Detti, G.~Z.~M. de~Lara, R.~Bruni, M.~Pranzo, F.~Sarnari, and G.~Vatti, ``A
  patient-specific approach for short-term epileptic seizures prediction
  through the analysis of eeg synchronization,'' \emph{IEEE Transactions on
  Biomedical Engineering}, vol.~66, no.~6, pp. 1494--1504, 2018.

\bibitem{zhang2019survey}
X.~Zhang, L.~Yao, X.~Wang, J.~Monaghan, and D.~Mcalpine, ``A survey on deep
  learning based brain computer interface: Recent advances and new frontiers,''
  \emph{arXiv preprint arXiv:1905.04149}, 2019.

\bibitem{altaf201516}
M.~A.~B. Altaf, C.~Zhang, and J.~Yoo, ``A 16-channel patient-specific seizure
  onset and termination detection soc with impedance-adaptive transcranial
  electrical stimulator,'' \emph{IEEE Journal of Solid-State Circuits},
  vol.~50, no.~11, pp. 2728--2740, 2015.

\bibitem{li2015feature}
J.~Li, Z.~Struzik, L.~Zhang, and A.~Cichocki, ``Feature learning from
  incomplete eeg with denoising autoencoder,'' \emph{Neurocomputing}, vol. 165,
  pp. 23--31, 2015.

\bibitem{goh2018spatio}
S.~K. Goh, H.~A. Abbass, K.~C. Tan, A.~Al-Mamun, N.~Thakor, A.~Bezerianos, and
  J.~Li, ``Spatio--spectral representation learning for electroencephalographic
  gait-pattern classification,'' \emph{IEEE Transactions on Neural Systems and
  Rehabilitation Engineering}, vol.~26, no.~9, pp. 1858--1867, 2018.

\bibitem{boashash2016automatic}
B.~Boashash and S.~Ouelha, ``Automatic signal abnormality detection using
  time-frequency features and machine learning: A newborn eeg seizure case
  study,'' \emph{Knowledge-Based Systems}, vol. 106, pp. 38--50, 2016.

\bibitem{thodoroff2016learning}
P.~Thodoroff, J.~Pineau, and A.~Lim, ``Learning robust features using deep
  learning for automatic seizure detection,'' in \emph{Machine learning for
  healthcare conference}, 2016, pp. 178--190.

\bibitem{arunkumar2018entropy}
N.~Arunkumar, K.~R. Kumar, and V.~Venkataraman, ``Entropy features for focal
  eeg and non focal eeg,'' \emph{Journal of computational science}, vol.~27,
  pp. 440--444, 2018.

\bibitem{yuan2019fusionatt}
Y.~Yuan and K.~Jia, ``Fusionatt: Deep fusional attention networks for
  multi-channel biomedical signals,'' \emph{Sensors}, vol.~19, no.~11, p. 2429,
  2019.

\bibitem{zhang2019adversarial}
X.~Zhang, L.~Yao, and F.~Yuan, ``Adversarial variational embedding for robust
  semi-supervised learning,'' in \emph{Proceedings of the 25th ACM SIGKDD
  International Conference on Knowledge Discovery \& Data Mining}, 2019, pp.
  139--147.

\bibitem{faust2015wavelet}
O.~Faust, U.~R. Acharya, H.~Adeli, and A.~Adeli, ``Wavelet-based eeg processing
  for computer-aided seizure detection and epilepsy diagnosis,''
  \emph{Seizure}, vol.~26, pp. 56--64, 2015.

\bibitem{bhattacharyya2017multivariate}
A.~Bhattacharyya and R.~B. Pachori, ``A multivariate approach for
  patient-specific eeg seizure detection using empirical wavelet transform,''
  \emph{IEEE Transactions on Biomedical Engineering}, vol.~64, no.~9, pp.
  2003--2015, 2017.

\bibitem{fan2018detecting}
M.~Fan and C.-A. Chou, ``Detecting abnormal pattern of epileptic seizures via
  temporal synchronization of eeg signals,'' \emph{IEEE Transactions on
  Biomedical Engineering}, vol.~66, no.~3, pp. 601--608, 2018.

\bibitem{schirrmeister2017deep}
R.~Schirrmeister, L.~Gemein, K.~Eggensperger, F.~Hutter, and T.~Ball, ``Deep
  learning with convolutional neural networks for decoding and visualization of
  eeg pathology,'' in \emph{SPMB}.\hskip 1em plus 0.5em minus 0.4em\relax IEEE,
  2017, pp. 1--7.

\bibitem{lin2016classification}
Q.~Lin, S.-q. Ye, X.-m. Huang, S.-y. Li, M.-z. Zhang, Y.~Xue, and W.-S. Chen,
  ``Classification of epileptic eeg signals with stacked sparse autoencoder
  based on deep learning,'' in \emph{International Conference on Intelligent
  Computing}.\hskip 1em plus 0.5em minus 0.4em\relax Springer, 2016, pp.
  802--810.

\bibitem{orosco2016patient}
L.~Orosco, A.~G. Correa, P.~Diez, and E.~Laciar, ``Patient non-specific
  algorithm for seizures detection in scalp eeg,'' \emph{Computers in biology
  and medicine}, vol.~71, pp. 128--134, 2016.

\bibitem{zhou2016attention}
P.~Zhou, W.~Shi, J.~Tian, Z.~Qi, B.~Li, H.~Hao, and B.~Xu, ``Attention-based
  bidirectional long short-term memory networks for relation classification,''
  in \emph{ACL (Volume 2: Short Papers)}, 2016, pp. 207--212.

\bibitem{amaral2013using}
T.~Amaral, L.~M. Silva, L.~A. Alexandre, C.~Kandaswamy, J.~M. Santos, and J.~M.
  de~S{\'a}, ``Using different cost functions to train stacked auto-encoders,''
  in \emph{2013 12th Mexican international conference on artificial
  intelligence}.\hskip 1em plus 0.5em minus 0.4em\relax IEEE, 2013, pp.
  114--120.

\bibitem{arjovsky2017wasserstein}
M.~Arjovsky, S.~Chintala, and L.~Bottou, ``Wasserstein gan,'' \emph{arXiv
  preprint arXiv:1701.07875}, 2017.

\bibitem{temko2011instantaneous}
A.~Temko, G.~Lightbody, E.~M. Thomas, G.~B. Boylan, and W.~Marnane,
  ``Instantaneous measure of eeg channel importance for improved
  patient-adaptive neonatal seizure detection,'' \emph{IEEE Transactions on
  Biomedical Engineering}, vol.~59, no.~3, pp. 717--727, 2011.

\bibitem{vaswani2017attention}
A.~Vaswani, N.~Shazeer, N.~Parmar, J.~Uszkoreit, L.~Jones, A.~N. Gomez,
  {\L}.~Kaiser, and I.~Polosukhin, ``Attention is all you need,'' in
  \emph{Advances in neural information processing systems}, 2017, pp.
  5998--6008.

\bibitem{zhang2017learning}
K.~Zhang, W.~Zuo, S.~Gu, and L.~Zhang, ``Learning deep cnn denoiser prior for
  image restoration,'' in \emph{CVPR}, 2017, pp. 3929--3938.

\bibitem{zhang2016mgnc}
Y.~Zhang, S.~Roller, and B.~Wallace, ``Mgnc-cnn: A simple approach to
  exploiting multiple word embeddings for sentence classification,''
  \emph{arXiv preprint arXiv:1603.00968}, 2016.

\bibitem{obeid2016temple}
I.~Obeid and J.~Picone, ``The temple university hospital eeg data corpus,''
  \emph{Frontiers in neuroscience}, vol.~10, p. 196, 2016.

\bibitem{ansari2019neonatal}
A.~H. Ansari, P.~J. Cherian, A.~Caicedo, G.~Naulaers, M.~De~Vos, and
  S.~Van~Huffel, ``Neonatal seizure detection using deep convolutional neural
  networks,'' \emph{International journal of neural systems}, p. 1850011, 2019.

\bibitem{kiral2018epileptic}
I.~Kiral-Kornek, S.~Roy, E.~Nurse, B.~Mashford, P.~Karoly, T.~Carroll,
  D.~Payne, S.~Saha, S.~Baldassano, T.~O'Brien \emph{et~al.}, ``Epileptic
  seizure prediction using big data and deep learning: toward a mobile
  system,'' \emph{EBioMedicine}, vol.~27, pp. 103--111, 2018.

\bibitem{lerche2013ion}
H.~Lerche, M.~Shah, H.~Beck, J.~Noebels, D.~Johnston, and A.~Vincent, ``Ion
  channels in genetic and acquired forms of epilepsy,'' \emph{The Journal of
  physiology}, vol. 591, no.~4, pp. 753--764, 2013.

\bibitem{zhang2019deep}
X.~Zhang, X.~Chen, L.~Yao, C.~Ge, and M.~Dong, ``Deep neural network
  hyperparameter optimization with orthogonal array tuning,'' in
  \emph{International Conference on Neural Information Processing}.\hskip 1em
  plus 0.5em minus 0.4em\relax Springer, 2019, pp. 287--295.

\end{thebibliography}
\end{document}